\documentclass[reqno,11pt]{amsart}
\usepackage{amssymb}
\usepackage{amsfonts}
\usepackage{amsmath}
\usepackage{graphicx}

\numberwithin{equation}{section}
\newtheorem{theorem}{Theorem}

\newtheorem{proposition}[theorem]{Proposition}
\newtheorem{remark}[theorem]{Remark}
\email{rflorese@gauss.mat.uson.mx}
\keywords{Schouten bracket, Lie transform method, tensor fields, Poisson tensor, Dirac bracket.}
\subjclass[2000]{34K33, 37J40, 53D17, 53Z05, 70H09}

\input{tcilatex}

\begin{document}
\title[Perturbations of Contravariant Antisymmetric Tensor Fields]{The Lie
Transform Method for Perturbations of Contravariant Antisymmetric Tensor
Fields and its Applications to Hamiltonian Dynamics}
\author{Rub\'en Flores-Espinoza}
\address{Depertament of Mathematics, University of Sonora\\
\hspace*{12pt}Blvd. Luis Encinas y Rosales, Centro\\
\hspace*{12pt}Hermosillo, M\'exico, 83000.}
\maketitle

\begin{abstract}
By means of the Schouten calculus for contravariant antisymmetric tensor
fields, we apply the Lie transform method to investigate smooth deformations
of tensor fields and, in particular, to perturbations of Hamiltonian systems
generated by deformations of the Poisson bracket. Using results by Karasev
and Vorobiev on the computation of Poisson cohomology we describe
infinitesimal generators for the Lie transformations. We give applications
to perturbed Euler equations on six dimensional Lie coalgebras and to
Hamiltonian systems on Poisson manifolds equipped with Dirac brackets.
\end{abstract}

\section{Introduction}

The method of Lie transforms is a powerful and general procedure to deal
with perturbed tensor fields of arbitrary type depending on a small
parameter. This method initially developed for vector fields by Deprit \cite%
{De}, Kamel \cite{K}, Hernard \cite{H}, and others, provides recursive
routines to transform the perturbation of the original system into another
one which has convenient y more tractable properties.

The basic idea is to use a parametrized family of near-identity
transformations defined by the evolution operator of a vector field
depending on the parameter. To obtain a transformed tensor field with some
desirable properties, the terms in the formal Taylor expansion on the
parameter of the vector field must satisfy a set of recursive homological
equations whose solvability depends on the properties we want for the
transformed tnsor field. Of course, in this case, the convergence of the
entire corresponding expansion is not guaranteed.

Here, we give an intrinsec presentation of the Lie transform method for
perturbations of contravariant antisymmetric tensor fields by using the
Schouten bracket calculus. This method is applied to more general
perturbations than the standard Hamiltonian ones for a fixed symplectic
structure. The possibility of the applicability of the Lie transform method
is related with the solvability of homological equations given in terms of
the Schouten bracket. In particular, in the case of Poisson tensors of
constant rank, an efficient method for solving homological equations was
suggested in \cite{KVo}. Here, we apply the Lie transform method to
perturbations of Hamiltonian systems generated by deformations of the
Poisson bracket, looking for families of near identity diffeomorphisms to
transform the perturbation of the Poisson bracket into a perturbation of the
Hamiltonian function for the unperturbed Hamiltonian system. When such a
transformation exists, on a second step we can apply some of the available
methods in the Hamiltonian perturbation theory (the averaging method, the
KAM theory, etcetera).

Finally we give two applications related with perturbed Euler equations on
six dimensional Lie coalgebras with perturbations of Hamiltonian systems on
Poisson manifolds equipped with Dirac brackets.\newline

\noindent\textbf{Acknowledgements}. The author is grateful to Guillermo
D\'avila-Rasc\'on and Yuri M. Vorobiev for fruitful discussions in the
preparation of this paper. This research was partially supported by CONACYT
under grant no. 55463.

\section{The Schouten Bracket and the Lie Transform}

Let $M$ a smooth manifold of dimension $m$. Denote by $V^{p}(M)$ the space
of contravariant antisymmetric $p$-tensor fields on $M$. In particular, $%
V^{0}(M)$ and $V^{1}(M)$ correspond to the space of smooth functions and
smooth vector fields on $M$, respectively. On the Grassman algebra $%
\displaystyle V(M) = \bigoplus_{i=0} ^{m}V^{i}(M)$ with the usual wedge
product $\wedge$, the Schouten bracket $[\negthinspace[ \,, ]\negthinspace]$
is defined as the unique local type operator extending the Lie derivative
and having the following properties for all $f \in V^{0}(M)$; $X,Y \in
V^{1}(M)$; $A \in {V^{p}(M)}$, $B \in {V^{q}(M)}$, $C \in {V^{r}(M)}$ \cite%
{Va}:%
\begin{align}
[\negthinspace[ \,, ]\negthinspace] :\ V^{p}(M) \times V^{q} (M) &
\rightarrow V^{p+q-1}(M)  \label{1.1} \\
[\negthinspace[ X, f ]\negthinspace] & = L_{X} f,  \label{1.2} \\
[\negthinspace[ X, Y ]\negthinspace] & = [X,Y],  \label{1.3} \\
[\negthinspace[ A, B ]\negthinspace] & = (-1)^{pq} [\negthinspace[ B, A ]%
\negthinspace],  \label{1.4} \\
[\negthinspace[ A, B \wedge{C} ]\negthinspace] & = [\negthinspace[ A, B ]%
\negthinspace] \wedge{C} + (-1)^{pq + q} B \wedge [\negthinspace[ A, C ]%
\negthinspace]   \label{1.5}
\end{align}

Moreover, the Schouten bracket satisfies the graded Jacobi identity:%
\begin{equation}
(-1)^{pr} [\negthinspace[{ [\negthinspace[ A, B ]\negthinspace], C }]%
\negthinspace] + (-1)^{qp} [\negthinspace[{ [\negthinspace[ B, C ]%
\negthinspace], A }]\negthinspace] + (-1)^{rq} [\negthinspace[{ [\negthinspace%
[ C, A ]\negthinspace], B }]\negthinspace] = 0.   \label{1.6}
\end{equation}
More details about the Schouten bracket calculus, can be found in \cite%
{Lichner, Va}.

Given a diffeomorphism $\gamma: M \rightarrow M$ and $A \in V^{p}(M),\ p
\geq 1$, \textit{the pull-back} $\gamma^{\ast} (A) \in V^{p}(M)$ of $A$
under $\gamma$, is defined by the usual formula,%
\begin{equation}
(\gamma^{\ast}A)(\mathrm{d} g_{1}, \ldots, \mathrm{d} g_{p})(x) = A \bigl( 
\mathrm{d} (g_{1} \circ \gamma^{-1}), \ldots, \mathrm{d}(g_{p} \circ
\gamma^{-1}) \bigr)(\gamma(x)),   \label{1.7}
\end{equation}
for all $g_{i} \in {V^{0}(M)}$, $i=1, 2, \ldots, p$. The pull-back $%
\gamma^{\ast}$ has the following properties 
\begin{align}
\gamma^{\ast}(A \wedge B) & = \gamma^{\ast} A \wedge \gamma^{\ast}B,
\label{1.8} \\
\gamma^{\ast} [\negthinspace[ A, B ]\negthinspace] & = [\negthinspace[
\gamma^{\ast} A, \gamma^{\ast}B ]\negthinspace].   \label{1.9}
\end{align}
for all $A \in V^{p}(M)$ and $B \in V^{q}(M)$.

Pick a $A_{0} \in V^{p}(M)$ and consider a deformation given by a smooth $%
\varepsilon$-family $A_{\varepsilon} \in V^{p}(M)$ with $A_{0} =
A\mid_{\varepsilon = 0}$. Consider an $\varepsilon$-family of
diffeomeorphisms $\gamma_{\varepsilon}$, $x \mapsto y =
\gamma_{\varepsilon}(x)$ defined by the solution of the Cauchy problem:%
\begin{align}
\frac{\mathrm{d}}{\mathrm{d} \varepsilon} \gamma_{\varepsilon} & =
X_{\varepsilon}(\gamma_{\varepsilon}),  \label{1.10} \\
\gamma_{0}(x) & = x, \qquad x \in M.  \notag
\end{align}
Here $X_{\varepsilon}$ is a smooth $\varepsilon$-depending vector field. It
follows from the ``variation of parameters formula''%
\begin{equation}
\frac{\mathrm{d}}{\mathrm{d} \varepsilon} \gamma_{\varepsilon}^{\ast}
A_{\varepsilon} = \gamma_{\varepsilon}^{\ast} [\negthinspace[
X_{\varepsilon} + \frac{\partial}{\partial\varepsilon}, A_{\varepsilon} ]%
\negthinspace],   \label{1.11}
\end{equation}
that if the vector field $X_{\varepsilon}$ satisfies the homological equation%
\begin{equation}
[\negthinspace[ X_{\varepsilon}, A_{\varepsilon} ]\negthinspace] = -\frac{%
\partial}{\partial\varepsilon} A_{\varepsilon},   \label{1.12}
\end{equation}
then $\gamma_{\varepsilon}^{\ast} A_{\varepsilon} = A_{0}$. Such deformation 
$A_{\varepsilon}$ of $A_{0}$ is called \textit{trivial}.

If the homological equation (\ref{1.12}) is solvable, then any $\varepsilon$%
-family of tensor fields of the form 
\begin{equation}
B_{\varepsilon} = [\negthinspace[ A_{\varepsilon}, C ]\negthinspace], 
\label{1.13}
\end{equation}
is tranformed by the near identity diffeomorphisms $\gamma_{\varepsilon}$
into the $\varepsilon$-family 
\begin{equation}
\gamma_{\varepsilon}^{\ast}B_{\varepsilon} = [\negthinspace[ A,
\gamma_{\varepsilon}^{\ast} C ]\negthinspace],   \label{1.14}
\end{equation}
where the deformation comes now from smooth changes of the $q$-tensor $C$.

The application of the above observation to deformations of tensor fields of
the form given before meet some difficulties. For example, the solvability
of the homological equation (\ref{1.12}) implies that all elements of the $%
\varepsilon$-parametrized smooth curve $A_{\varepsilon} \in V^{p}(M)$ share
with $A_{0}$ the same topological or geometric properties. In practice, this
conditions are not satisfied globally and hence, the solution of the
homological equation (\ref{1.12}) does not exists or it exists only on some
restricted domains. To deal with those problems, in the frame of the
averaging theory we use the construction of approximate solutions for the
homological equation with an error up to some given power of the small
parameter $\varepsilon$.

In fact, if the Taylor expansion of $A_{\varepsilon}$ and $X_{\varepsilon}$
is given by 
\begin{align}
A_{\varepsilon} & = \sum_{i\geq0}\frac{\varepsilon^{i}}{i!} A_{i},
\label{1.15} \\
X_{\varepsilon} & = \sum_{i\geq0}\frac{\varepsilon^{i}}{i!} X_{i}, 
\label{1.16}
\end{align}
and we substitute these expresions in the homological equation (\ref{1.12}),
we obtain the recursive set of homological equations for vector fields $X_{0}
$, $X_{1}$, \ldots, $X_{k}$, \ldots, 
\begin{align}
[\negthinspace[ X_{0}, A_{0} ]\negthinspace] & = - A_{1}  \label{1.17} \\
[\negthinspace[ X_{1}, A_{0} ]\negthinspace] & = - [\negthinspace[ X_{0},
A_{1} ]\negthinspace] - A_{2}  \label{1.18} \\
[\negthinspace[ X_{2}, A_{0} ]\negthinspace] & = - 2 [\negthinspace[ X_{1},
A_{1} ]\negthinspace] - [\negthinspace[ X_{0}, A_{2} ]\negthinspace] - A_{3}
\notag \\
& \vdots  \notag \\
[\negthinspace[ X_{k}, A_{0} ]\negthinspace] & = - \sum_{i=1}^{k} \binom{k}{i%
} [\negthinspace[ X_{k-i}, A_{i} ]\negthinspace] - A_{k+1}.  \label{1.20} \\
& \vdots  \notag
\end{align}
Note that these equations can be written in the recursive form $[%
\negthinspace[ X_{k}, A_{0} ]\negthinspace] = F(A_{1}, \ldots, A_{k+1})$.
From the discusion above we derive the following result.

\begin{proposition}
Let $A_{\varepsilon}\in V^{p}(M)$ a smooth deformation of the contravariant
antisymmetric $p$-tensor field $A_{0}$ and $X_{0},X_{1}, \ldots, X_{k},
\ldots$ vector fields satisfying the set of recursive homological equations {%
\upshape (\ref{1.17})}, {\upshape (\ref{1.18})} and {\upshape (\ref{1.20})}
in a domain $D$. Then, for each tensor field $C \in V^{q}(M)$, the $%
\varepsilon$-family of vector fields $[\negthinspace[ A_{\varepsilon}, C ]%
\negthinspace]$ is transformed up to $O(\varepsilon^{k})$ into the $%
\varepsilon$-family of tensor fields $[\negthinspace[ A_{0}, C_{\varepsilon} 
]\negthinspace]$ where 
\begin{equation}
C_{\varepsilon} = C + \sum_{i=1}^{k} \frac{\varepsilon^{i}}{i!} [%
\negthinspace[ X_{i-1}, C ]\negthinspace] + O(\varepsilon^{k}). 
\label{1.21}
\end{equation}
\end{proposition}

\section{Solvability of the homological equation for perturbed Poisson
tensors}

In order to apply the above techniques to Hamiltonian dynamics on Poisson
manifolds, we consider, in the sequel, perturbations of the form $%
A_{\varepsilon} = [\negthinspace[ \Psi_{\varepsilon}, H ]\negthinspace]$,
where $\Psi_{\varepsilon} \in V^{2}(M)$ and $H \in V^{0}(M).$ In this case,
the set of recursive equations (\ref{1.17}), (\ref{1.18}) and (\ref{1.20})
take the special form, 
\begin{equation}
[\negthinspace[ X, \Psi ]\negthinspace] = \Phi,   \label{2.1}
\end{equation}
for tensors fields $\Psi,\Phi \in V^{2}(M)$. Such a class of homological
equations has been studied in \cite{KVo} for the case when $\Psi$ is a
Poisson tensor of constant rank and its symplectic foliation is a fibration.
We recall here some basic facts about Poisson manifolds and the results
given in \cite{KVo}.

A \textit{Poisson manifold} is a pair $(M,\Psi)$, where $\Psi \in V^{2}(M)$
and $[\negthinspace[ \Psi, \Psi ]\negthinspace] = 0$. The distinguished
quantities in a Poisson manifold are: the \textit{Casimir functions} $k \in
V^{0}(M)$ with $[\negthinspace[ \Psi, k ]\negthinspace] = 0$; the \textit{%
Hamiltonian vector fields} defined as $X_{f} = [\negthinspace[ \Psi, f ]%
\negthinspace]$ for each $f \in V^{0}(M)$. The \textit{Poisson vector fields}
$X \in V^{1}(M)$ or \textit{infinitesimal automorphisms} are defined as $[%
\negthinspace[ \Psi, X ]\negthinspace] = 0$. On a Poisson manifold, the
operation between each pair of functions $f, g \in V^{0}(M)$ given by%
\begin{equation}
\left\{ f, g \right\} = [\negthinspace[{ [\negthinspace[ \Psi, f ]%
\negthinspace], g }]\negthinspace],   \label{2.2}
\end{equation}
is called the \textit{Poisson bracket} on $M$.

At each point $x\in M,$ the dimension of the subspace in $T_{x}M$ generated
by the vectors $[\negthinspace[ \Psi, f ]\negthinspace] (x)$ is even and
equals to the rank of the $2$-tensor field $\Psi$. If the rank of $\Psi$ is
constant on $M$, then the Hamiltonian vector fields define an integrable
distribution $\mathcal{D}$ in the sense of Frobenius. Each leaf $\mathcal{L}$
of the regular foliation is a symplectic manifold whose symplectic form is a 
$2$-form $\omega_{\mathcal{L}}$ given by the formula%
\begin{equation}
\omega_{\mathcal{L}}(X_{f},X_{g}) = \left\{ f, g \right\}.   \label{2.3}
\end{equation}
This means that field $\Psi$ defines on each symplectic leaf $\mathcal{L}$,
an isomorphism between $\Lambda^{1}(\mathcal{L})$ and $V^{1}(\mathcal{L})$.
For a mmore comprehensive study of Poisson manifolds see \cite{KMs, Lichner,
Va}.

We assume that $\mathrm{rank}\, \Psi = 2r$ and the symplectic foliation is a
fibration and the space of leaves is an open domain in $\mathbb{R}^{m-2r}$.
Then, the symplectic foliation is given by the level sets of $m-2r$
independent Casimir functions $k_{1}, \ldots, k_{m-2r}$.

Recall that tensors fields on $M$, tangent to symplectic leaves are called
vertical. A smooth field of $k$-forms along the leaves is called a vertical $%
k$-form on $M$.

We have a correspondance $\Psi^{\sharp}$ between vertical forms and vertical
contravariant antisymmetric tensor fields on $M$ given by the expresion%
\begin{equation}
(\Psi^{\sharp} \alpha) (\mathrm{d} f_{1}, \ldots, \mathrm{d} f_{r}) = \alpha %
\bigl(X_{f_{1}}, \ldots, X_{f_{r}} \bigr), \qquad f_{1}, \ldots, f_{r} \in
V^{0}(M).   \label{2.4}
\end{equation}
where $\alpha$ is a smooth section of $r$-forms of the bundle $\pi$.

A necessary condition for the solvability of (\ref{2.1}) is 
\begin{equation}
[\negthinspace[ \Psi, \Phi ]\negthinspace] = 0,   \label{2.5}
\end{equation}
This means that the tensor field $\Phi$ is a $2$-cocycle of the
Poisson-Lichnerowicz coboundary operator $\delta_{\Psi}$ on the Grassmann
algebra of contravariant antisymmetric tensor fields defined by $%
\delta_{\Psi}(A) = [\negthinspace[ \Psi, A ]\negthinspace]$ for $A \in
V^{p}(M)$, $p = 0, 1, \ldots, \dim M$. Moreover, the solvability of (\ref%
{2.1}) is equivalent to the condition that $\Phi$ is a $2$-coboundary for
the operator $\delta_{\Psi}$.

\begin{remark}
In terms of the coboundary operator $\delta_{\Psi}$, equation {\upshape (\ref%
{2.1})} for an infinitesimal generator $X$ of the Lie transformation {%
\upshape (\ref{1.17})--(\ref{1.20})} is written as $\delta_{\Psi} X = \Phi$.
On the other hand, in terms of the Lie derivative $L_{X}$ along $X$, this
equation takes the form $L_{X}\Psi=\Phi$. In the averaging method and normal
forms \cite{AKN, De}, such a type of equations are usually caled homological
equations.
\end{remark}

If a $2$-tensor field $\Phi$ is a $2$-coboundary of $D$, we have 
\begin{equation}
[\negthinspace[ \Psi, X ]\negthinspace] (\mathrm{d} k_{i}) = \Phi(k_{i}),
\qquad i = 1, \ldots, m-2r,   \label{2.6}
\end{equation}
or 
\begin{equation}
\Psi(L_{X}k_{i}) = \Phi(k_{i}), \qquad i = 1, \ldots, m-2r   \label{2.7}
\end{equation}
i.e., $\Phi(k_{i})$, $i = 1, \ldots, r$ have to be Hamiltonian vector fields
with respect to the Poisson tensor $\Psi$. This last condition is also
sufficient. To see that, suppose $\Phi(k_{i}) = \Psi(h_{i})$, $i = 1,
\ldots, m-2r$ and consider vector fields $V_{i}$ dual to the Casimir
functions, i.e., $L_{V_{i}}k_{j} = \delta_{i}^{j}$, $i,j = 1, \ldots, m-2r$.
Then 
\begin{equation*}
(\Phi - \sum_{i=1}^{r} [\negthinspace[ \Psi, h_{i} V_{i} ]\negthinspace] )
(k_{j}) = 0, 
\end{equation*}
and such tensor becomes a vertical $2$-tensor in the bundle $\pi$. Taking
into account that $\Psi^{\sharp}$ is an isomorphism, we can assure the
existence of a smooth section $\alpha$ of 2-forms on $\pi$ such that 
\begin{equation}
\Psi^{\sharp} \alpha = \Phi - \sum_{i=1}^{r} [\negthinspace[ \Psi,
h_{i}V_{i} ]\negthinspace].   \label{2.8}
\end{equation}
Moreover, by condition (\ref{2.5}) we have%
\begin{equation}
0 = [\negthinspace[ \Psi, \Psi^{\sharp}\alpha ]\negthinspace] =
\Psi^{\sharp} \mathrm{d} \alpha,   \label{2.9}
\end{equation}
and $\alpha$ is a closed vertical $2$-forms. If we suppose that each
symplectic leaf is simply connected, then $\alpha = d \beta$ on symplectic
leaves, $\Psi^{\sharp} \alpha = [\negthinspace[ \Psi, \Psi^{\sharp} \beta ]%
\negthinspace]$ and the vector field on $M$%
\begin{equation}
X = \Psi^{\sharp} \beta + \sum_{i=1}^{r} h_{i} V_{i} + \left( \text{Poisson
vector field} \right),   \label{2.10}
\end{equation}
satisfies the homological equation (\ref{2.5}).

We now recall the following theorem given in \cite{KVo}.

\begin{proposition}
\label{pro:karvor}If the Poisson tensor of a Poisson manifold $(M,\Psi)$ has
constant rank $2r$ and its symplectic foliation is a fibration by simply
connected leaves, then the homological equation {\upshape (\ref{2.5})} is
solvable if and only if the vector fields $\Phi(k_{j})$ are Hamiltonian
vector fields with respect to $\Psi$ for $j=1, \ldots, m-2r$. In this case,
all solutions of {\upshape (\ref{2.4})} are of the form {\upshape (\ref{2.10}%
)}.
\end{proposition}

\begin{remark}
Under the assumptions of Proposition {\upshape \ref{pro:karvor}}, for any
element $\Psi_{\varepsilon}$ of a smooth $\varepsilon$-family of Poisson
tensors, $[\negthinspace[ \Psi_{\varepsilon}, \Psi_{\varepsilon} ]%
\negthinspace] = 0$, and $k_{\varepsilon}$ a smooth $\varepsilon$-dependent
Casimir function, $[\negthinspace[ \Psi_{\varepsilon}, k_{\varepsilon} ]%
\negthinspace] = 0$, then 
\begin{equation*}
[\negthinspace[ \frac{\mathrm{d}}{\mathrm{d} \varepsilon} \Psi_{%
\varepsilon}, k_{\varepsilon} ]\negthinspace] = - [\negthinspace[
\Psi_{\varepsilon}, \frac{\mathrm{d}}{\mathrm{d} \varepsilon}k_{\varepsilon} 
]\negthinspace], 
\end{equation*}
and the homological equation {\upshape (\ref{1.12})} is solvable over an
open domain in $\mathbb{R}^{2r}$.
\end{remark}

\section{Perturbations of Hamiltonian Systems Generated by Deformations of
Poisson Brackets}

On the Poisson manifold $(M,\Psi)$, let us consider a Hamiltonian system%
\begin{equation}
\dot{y} = [\negthinspace[ \Psi, H ]\negthinspace] (y), \qquad y \in M, 
\label{3.1}
\end{equation}
and a perturbed system of the form%
\begin{equation}
\dot{y} = [\negthinspace[ \Psi_{\varepsilon}, H ]\negthinspace] (y), \qquad
y \in M,   \label{3.2}
\end{equation}
where $\Psi_{\varepsilon}$ is an smooth $\varepsilon$-family of $2$-tensors
fields $\Psi_{\varepsilon}$ with $\Psi_{0} = \Psi$.

In general, the perturbation vector field%
\begin{equation*}
X_{\mathrm{pert}} = [\negthinspace[ \Psi_{\varepsilon} - \Psi, H ]%
\negthinspace] (y), \qquad y \in M, 
\end{equation*}
associated to (\ref{3.2}) is not a Hamiltonian vector field with respect to
the initial Poisson structure $\Psi$. In fact, $X_{\mathrm{pert}}$ can be
transversal to the symplectic leaves of $\Psi$ as we will see in the
applications given below.

Finding conditions under which one can transform the perturbed part of the
Hamiltonian system (\ref{3.2}) generated by deformations of the Poisson
bracket into another perturbed system where the perturbation comes from
smooth changes in the Hamiltonian function $H$, is a relevant question for
the applications. When such transformations exist, we say that the
transformed system is in \textit{normal form}. Therefore, if system (\ref%
{3.2}) can be taken into normal form, one can apply standar methods in the
perturbation theory for Hamiltonian systems, see for example \cite{Tresch}.

As an illustration of the above ideas we present two examples.

\subsection{Perturbations of Euler systems on Lie-coalgebras}

On the $6$-dimensional Euclidean space $\mathbb{R}^{6}= \mathbb{R}_{y}^{3}
\times \mathbb{R}_{z}^{3}$ with coordinates $y = (y_{1},y_{2},y_{3})$, $z=
(z_{1},z_{2},z_{3})$ consider the diagonal matrix 
\begin{equation}
\eta = \mathrm{diag}(\eta_{1},\eta_{2},\eta_{3}) = 
\begin{bmatrix}
\eta_{1} & 0 & 0 \\ 
0 & \eta_{2} & 0 \\ 
0 & 0 & \eta_{3}%
\end{bmatrix}%
,   \label{3.3}
\end{equation}
and the $\varepsilon-$family of Poisson tensors that have relevance in the
rigid body motion \cite{Bo} 
\begin{align}
\Psi_{\eta,\varepsilon} (\mathrm{d} f,\mathrm{d} g) & = (\eta(y) \times
\nabla_{y} f + \eta(z) \times \nabla_{z}f) \cdot \nabla_{y} g  \notag \\
& + (\eta(z) \times \nabla_{y} f + \varepsilon \eta(z) \times \nabla_{z}f)
\cdot \nabla_{z} g.  \label{3.4}
\end{align}
for some functions $f,g \in V^{0}( \mathbb{R}^{6})$. The $2$-tensor field (%
\ref{3.4}) is a linear Poisson tensor and arbitrary $\eta$ and can be
written in the form 
\begin{equation}  \label{ejem1rb}
\Psi_{\eta,\varepsilon} = \Psi_{\eta,0} + \varepsilon \, \Phi_{\eta}
\end{equation}
with 
\begin{equation*}
\Phi_{\eta}(\mathrm{d} f, \mathrm{d} g) = \bigl( \eta(y) \times \nabla_{z}f %
\bigr) \cdot \nabla_{z}g. 
\end{equation*}
For the given values of $\eta$ and $\varepsilon$, below, there corresponds
the Lie-Poisson bracket on the coalgebra $\mathfrak{g}^{\ast}$ of the
following six dimensional Lie algebras:

\begin{center}
\begin{tabular}{ccc}
$\eta$ & $\varepsilon$ & Lie-Poisson bracket on $\mathfrak{g}^{\ast}$ \\ 
\hline
&  &  \\ 
$\eta = (1,1,1)$ & $\varepsilon > 0$, & $\mathfrak{so}(4)^{\ast}$ \\ 
&  &  \\ 
$\eta = (1,1,1)$ & $\varepsilon = 0$, & $\mathfrak{e}(3)^{\ast}$ \\ 
&  &  \\ 
$\eta = (1,1,-1)$ & $\varepsilon > 0$, & $\mathfrak{so}(2,2)^{\ast}$ \\ 
&  &  \\ 
$\eta = (1,1,-1)$ & $\varepsilon = 0$, & $\mathfrak{l}(3)^{\ast}$ \\ 
&  &  \\ 
$\eta = (1,1,1) $ & $\varepsilon < 0$, & $\mathfrak{so}(3,1)^{\ast}$ \\ 
&  &  \\ \hline
\end{tabular}
\end{center}

\vspace{10pt}

For the different cases above, the Casimir functions are%
\begin{equation}  \label{3.5}
k_{\eta,\varepsilon}^{1}(y,z) = \eta(y) \cdot z, \qquad
k_{\eta,\varepsilon}^{2}(y,z) = \frac{1}{2} \varepsilon \eta(y) \cdot y + 
\frac{1}{2} \eta(z) \cdot z,
\end{equation}
and its regular symplectic leaves for $\varepsilon>0$ are the regular
coadjoint orbits on $\mathfrak{g}^{\ast}$ and diffeoemorphic to to $\Sigma
\times \Sigma$ where%
\begin{equation}
\Sigma = \left\{ \xi \in \mathbb{R}^{3} \; : \; \eta(\xi) \cdot \xi = 
\mathrm{const} \right\}.   \label{3.6}
\end{equation}
In the limit $\varepsilon \rightarrow 0$, the regular symplectic leaves of $%
\Psi_{\eta,0}$ are diffeomorphic to $T^{\ast} \Sigma$. If $\Sigma$ is a $2$%
-dimensional sphere or a cylinder, then the symplectic leaves of $%
\Psi_{\varepsilon}$ and $\Psi_{0}$ have different topological structure and
the homological equation (\ref{1.12}) has no global solutions.

For a each diagonal matrix $\eta$ and smooth function $H(y,z) \in V^{0}(%
\mathbb{R}^{6})$, let us consider the Hamiltonian system on $(\mathbb{R}%
^{6}, \Psi_{\eta,0})$%
\begin{align}
\dot{y} & = \eta(y) \times \nabla_{y} H + \eta(z) \times \nabla_{z} H,
\label{3.7} \\
\dot{z} & = \eta(z) \times \nabla_{y} H,   \label{3.8}
\end{align}
and the $\varepsilon$-perturbed system of the form 
\begin{align}
\dot{y} & = \eta(y) \times \nabla_{y} H + \eta(z) \times \nabla_{z} H
\label{3.9} \\
\dot{z} & = \eta(z) \times \nabla_{y} H + \varepsilon \eta (y)
\times\nabla_{z} H   \label{3.10}
\end{align}
where $\varepsilon \in (-a,a)$. The Hamiltonian systems (\ref{3.7}), (\ref%
{3.8}) and (\ref{3.9}), (\ref{3.10}) are usually called Euler systems.

The vector fields associated to (\ref{3.7}), (\ref{3.8}) and (\ref{3.9}), (%
\ref{3.10}) take the form $[\negthinspace[ \Psi_{\eta,0}, H ]\negthinspace]$
and $[\negthinspace[ \Psi_{\eta,\varepsilon}, H ]\negthinspace]$,
respectively. The homological equation (\ref{1.12}) has a solution $%
X_{\varepsilon}$ on the domain 
\begin{equation*}
\mathcal{N} = \left\{ (y,z) \in \mathbb{R}^{6} \; : \; (\eta(y)
\times\eta(z) \cdot(y\times z) \neq 0 \right\}, 
\end{equation*}
given by 
\begin{equation}
X_{\varepsilon} = \frac{\eta(y) \cdot y}{2 \, \eta(y) \times \eta(z) \cdot
(y\times z)} \eta(y) \times(z\times y) \frac{\partial}{\partial z}, 
\label{3.11}
\end{equation}
with flow%
\begin{equation}  \label{3.12}
\gamma_{\varepsilon}(y,z) = \left(y,\alpha z + (1 - \alpha) \frac{\eta(y)
\cdot z} {\eta(y) \cdot y} y\right).
\end{equation}
Here, 
\begin{equation}  \label{3.13}
\alpha = \left[ 1 + \frac{\varepsilon(\eta(y) \cdot y)^{2}}{(\eta(y) \times
\eta(z) \cdot (y\times z)} \right]^{\frac{1}{2}},
\end{equation}
and the perturbation in system (\ref{3.9}), (\ref{3.10}) is transformed
under $\gamma_{\varepsilon}$ into a Hamiltonian perturbation relative to the
structure $\Psi_{\eta,0}$, which takes the form%
\begin{align}
\dot{y} & = \eta(y) \times \nabla_{y} H_{\varepsilon} + \eta(z) \times
\nabla_{z} H_{\varepsilon},  \label{3.14} \\
\dot{z} & = \eta(z) \times \nabla_{y} H_{\varepsilon},   \label{3.15}
\end{align}
where%
\begin{equation}  \label{3.16}
H_{\varepsilon}(y,z) = H \left(y,\alpha z + (1 - \alpha) \frac{\eta(y) \cdot
z}{\eta(y) \cdot y} y \right).
\end{equation}

If instead of solving the equation (\ref{1.12}) and thinking of $%
\Psi_{\eta,\varepsilon}$ as an exact homotopy of $\Psi_{\eta,0}$, we look
for an homotopy up to $O(\varepsilon^{k})$, then we can show that the
conditions for their solvability only requires that unperturbed tensor $%
\Psi_{\eta,0}$ to satisfy the conditions of Proposition \ref{pro:karvor}. In
our case, $\Psi_{\eta,0}$ has maximum rank in the set $\mathcal{D} = \left\{
(y,z)\in \mathbb{R}^{6} \; : \; \eta (z) \neq 0 \right\}$ and the symplectic
foliation is a trivial fibration over an open subset of $\mathbb{R}^{2}$,
hence each equation in (\ref{1.20}) has a solution on $\mathcal{D}$. In
particular,%
\begin{equation}
X_{0} = \frac{1}{2\eta(z) \cdot z} \eta(y) \times(z\times y) \frac{\partial}{%
\partial z}.   \label{3.28}
\end{equation}
This example was considered in \cite{KVoFl}.

\subsection{Deformations of Dirac brackets}

Let $M$ be a smooth manifold and $w_{\varepsilon}$ a smooth $\varepsilon$%
-family of nondegenerate closed $2$-forms (symplectic structures) on $M$ for 
$\varepsilon \in{(-a,a)}$. Denote by $\Psi_{\varepsilon}$ the nondegenerate
Poisson structure on $M$ associated to $\mathit{w_{\varepsilon}}$. Suppose
we have a set of functionally independent functions $\mathcal{A}%
_{\varepsilon}^{1}, \ldots, \mathcal{A}_{\varepsilon}^{r} \in V^{0}(M)$,
smoothly depending on $\varepsilon$. Denote 
\begin{equation*}
\Delta(\varepsilon) = ((\Delta^{ij})) \equiv \bigl( \bigl( \lbrack 
\negthinspace[{ [\negthinspace[ \Psi_{\varepsilon}, \mathcal{A}%
_{\varepsilon}^{i} ]\negthinspace], \mathcal{A}_{\varepsilon}^{j} }]%
\negthinspace] \bigr)\bigr), 
\end{equation*}
and assume that condition $\det(\Delta) \neq {0}$ holds everywhere on $M$,
for all $\varepsilon$. For every $\varepsilon \in (-a,a)$, define the 
\textit{Dirac tensor} by the standard formula (see, for example \cite{KMs}), 
\begin{equation}
\Psi_{\varepsilon}^{\mathrm{DIR}} = \Psi_{\varepsilon} + \frac{1}{2}
\sum_{(i,j)} \Delta_{ij}(\varepsilon) [\negthinspace[ \Psi_{\varepsilon}, 
\mathcal{A}_{\varepsilon}^{i} ]\negthinspace] \wedge [\negthinspace[
\Psi_{\varepsilon}, \mathcal{A}_{\varepsilon}^{j} ]\negthinspace], 
\label{3.17}
\end{equation}
where $\Delta^{is} \Delta_{sj} = \delta_{j}^{i}$. Let us consider the family
of Poisson brackets (\ref{3.17}) as a deformation of the ``unperturbed''
Dirac structure $\Psi_{0}^{\mathrm{DIR}}$. The symplectic leaves $\mathcal{S}%
_{\varepsilon}$ of (\ref{3.17}) coincide with the level sets of the
constraint functions $\mathcal{A}_{\varepsilon}^{i}$, 
\begin{equation}
\mathcal{S}_{\varepsilon} = \left\{ \mathcal{A}_{\varepsilon}^{1} = \mathrm{%
const},\ldots, \mathcal{A}_{\varepsilon}^{r} = \mathrm{const}\right\}. 
\label{3.18}
\end{equation}
In this case, the rank is constant and the forms $\displaystyle \dot{w}%
_{\varepsilon} = \frac{\mathrm{d} w_{\varepsilon}}{\mathrm{d} \varepsilon}$
are closed for every value of the parameter $\varepsilon$.

Now, suppose that there exists a smooth family of $1$-forms $%
\theta_{\varepsilon}$ on $M$ such that 
\begin{equation}
\dot{w}_{\varepsilon} = \mathrm{d} \theta_{\varepsilon} \quad \mathrm{mod} (%
\mathcal{F}_{\mathcal{S}}^{2}(M)), \qquad \varepsilon \in (-a,a). 
\label{3.19}
\end{equation}
Here $\mathcal{F}_{\mathcal{S}}^{2}(M)$ is the subspace of the $2$-forms on $%
M$ vanishing along the leaves $\mathcal{S}_{\varepsilon}$. So, in particular
the $2$-forms $\dot{w}_{\varepsilon}$ are exact on each symplectic leaf $%
\mathcal{S}_{\varepsilon}$. Notice that if in addition to this condition the
symplectic foliation (\ref{3.18}) is a trivial fibration by simply connected
leaves, then (\ref{3.19}) holds. As is known \cite{FlVo2}, Dirac structures
possesses a maximun set of Poisson vector fields which are transversal to
the symplectic leaves. In fact, the vector fields 
\begin{equation}
Z_{i}^{\varepsilon} = \sum_{j=1}^{r} \Delta_{ji} [\negthinspace[
\Psi_{\varepsilon}, \mathcal{A}_{\varepsilon}^{i} ]\negthinspace], \qquad i
= 1, \ldots, r,   \label{3.20}
\end{equation}
constitute a set of independent transversal Poisson vector fields. The
Poisson manifolds having such property are called transversally constant 
\cite{Va} or transversally maximal \cite{FlVo2}. Then, the homological
equation (\ref{1.12}) is solvable on $M$ for all $\varepsilon \in (-a,a)$
and the corresponding time-dependent solution can be represented by 
\begin{equation}  \label{3.21}
X_{\varepsilon} = \sum_{(i,j)} \Delta_{ij} \dot{\mathcal{A}}%
_{\varepsilon}^{j} [\negthinspace[ \Psi_{\varepsilon}, \mathcal{A}%
_{\varepsilon}^{i} ]\negthinspace] - \Psi_{\varepsilon}^{\mathrm{DIR}}
\theta_{\varepsilon},
\end{equation}
where $\dot{\mathcal{A}}_{\varepsilon}^{j} = \frac{\mathrm{d} \mathcal{A}%
_{\varepsilon}^{j}}{\mathrm{d} \varepsilon}$.

\vspace{20pt}

\end{document}